\documentclass[aps,prl,12pt,showkeys]{revtex4}

\usepackage{graphicx}

\begin{document}
\title{\bf On the vortex interpretation of Schr\"{o}dinger's wave equation}
\author{J. Orlin Grabbe}
\email[Email: ]{quantum@orlingrabbe.com}
\date{October 8, 2005}

\begin{abstract}
There has been a recent tendency to apply Schr\"{o}dinger's wave equation to macroscopic domains, from Bose-Einstein condensates in neutron stars to planetary orbits.  In these applications a hydrodynamical interpretation, involving vortices in some 'fluid' medium is often given. The vortex picture appears surprising in light of more traditional interpretations of Schr\"{o}dinger's equation, and indeed often appears to rely on ad hoc analogies.  The purpose of this letter is to examine the vortex hypothesis in light of a simple, transparent mathematical framework. We find that Schr\"{o}dinger's equation implies waves can be vortices also for a class of wave functions. Vortices occur in pairs that perform a sort of quantum computation by collapsing into either 0 or 1. Vortices collapsing to 0 ('0-vortices') are longer-lived, and their ratio to 1-vortices is discussed.

\end{abstract}


\maketitle

\subsection{Introduction}

Can a wave be a vortex?  The circular motion of vortices is seen in spiral galaxies, hurricanes, and quantum fluids.
Are all these, in some sense, the same thing?  Is it legitmate to apply Schr\"{o}dinger's wave equation to all of
them?  Is it acceptable to establish an analogy between Schr\"{o}dinger's equation and a hydrodynamic or similar equation in order to invoke vortices, and then to return to Schr\"{o}dinger's equation to invoke quantized observables of some type---whether energy levels or planetary orbits?  The Bohr-Sommerfeld model of quantized energy levels of electrons around the nucleus, or quantized angular momentum, could be visualized as little planets revolving around a nucleus-sun.  Is it now legitimate to invoke quantum theory to describe planetary orbits?  Wasn't Bohr's theory superseded by the \emph{new} quantum theory of Heisenberg and Schr\"{o}dinger?     

This letter addresses a single primitive aspect of these various questions:  Can Schr\"{o}dinger's wave equation imply vortices without ad hoc appeals to other areas of physics? If so, the goal is to demonstrate this in a simple, transparent methematical framework. We want to know if a vortex can arise from a wave in Schr\"{o}dinger's wave equation.

\subsection{The wave function as the complex power $z^c$}

Consider the wave function $\psi = z^c$, where $z > 0$ is real, and $c = x + iy$ is complex.  We may expand $\psi$, a complex number $\psi = u + iv$, as follows:
\begin{equation}
\psi = u+iv = z^c = e^{c \mbox{ }ln \mbox{ }z} = e^{x \mbox{ }ln\mbox{ }z + iy\mbox{ }ln\mbox{ }z} = z^x [cos(y\mbox{ ln }z) + i\mbox{ } sin(y\mbox{ ln }z)] .
\end{equation}
For $\psi$ to be analytic, we need it to satisfy the Cauchy-Riemann equations:
\begin{eqnarray}
\frac{\partial u}{\partial x} = \frac{\partial v}{\partial y} \\
\frac{\partial u}{\partial y} = -\frac{\partial v}{\partial x} 
\end{eqnarray}
where
\begin{eqnarray}
u = z^x cos(y\mbox{ ln }z)\\
v = z^x sin(y\mbox{ ln }z) .
\end{eqnarray}
We have
\begin{eqnarray}
\frac{\partial u}{\partial x} = z^x \mbox{ ln }z\mbox{ }cos(y\mbox{ ln }z)\\
\frac{\partial v}{\partial y} = z^x \mbox{ ln }z\mbox{ }cos(y\mbox{ ln }z)\\
\frac{\partial u}{\partial y} = -z^x \mbox{ ln }z\mbox{ }sin(y\mbox{ ln }z)\\
\frac{\partial v}{\partial x} = z^x \mbox{ ln }z\mbox{ }sin(y\mbox{ ln }z).
\end{eqnarray}
Hence the Cauchy-Riemann conditions are satisfied, and we may take the complex derivative
\begin{equation}
\frac{d\psi}{d c} = \frac{\partial u}{\partial x} + i \frac{\partial v}{\partial x} = (\mbox{ln }z)\mbox{ }z^c = (\mbox{ln }z)\mbox{ } \psi .
\end{equation}
That is, the derivative of $\psi$ with respect to $c$ is equal to $\psi$ multiplied by $\mbox{ln }z$.  Taking the derivative simply contracts or stretches the length of $\psi$ according to whether the absolute value of $\mbox{ln }z$ is less than or greater than 1 (and changes the direction of $\psi$ if $\mbox{ln }z$ is negative). 

Simularly we have
\begin{equation}
\frac{d^2\psi}{d c^2} = (\mbox{ln }z)^2\mbox{ } \psi
\end{equation}
The second derivative of $\psi$ with respect to $c$ simply multiplies $\psi$ by $(\mbox{ln }z)^2$.  That is,
$(\mbox{ln }z)^2$ is the eigenvalue of the differential operator $\frac{d^2}{d c^2}$ with respect to the
eigenvector $\psi$.

Because $\psi = u+iv$ is analytic, $u$ and $v$ satisfy Laplace's equation
\begin{equation}
\frac{\partial^2 u}{\partial x^2} + \frac{\partial^2 u}{\partial y^2} = 0 = \frac{\partial^2 v}{\partial x^2} + \frac{\partial^2 v}{\partial y^2} .
\end{equation}
Agaih, because $\psi (c)$ is analytic, we have from Cauchy's theorem that for a simple closed curve $C$ that is sufficiently smooth
\begin{equation}
\oint_C \psi(c) dc = 0
\end{equation}
and the additional result that at a point $c = a$ inside C
\begin{equation}
\psi(a) = \frac{1}{2\pi i} \oint_C \frac{\psi(c)}{c-a} dc .
\end{equation}

Note that since $\psi^*\psi = z^{2x}$, some restrictions are necessary in order for $\psi$ to be normalizable.
The integral $\int_{-\infty}^{\infty} e^{2x\mbox{ }ln\mbox{ }z} dx$ converges provided $z<1, x>0$ or else $z>1, x< 0$.
Otherwise we must restrict $x$ to $x \in D$, where $D$ is some finite domain.

We now let $z = z(r_x,r_y,t)$ where $r_x$ and $r_y$ are position coordinates, and $t$ is time.  (Note that we have already used $x$ and $y$ as components of the complex $c$---they are not position coordinates.) Thus we have
\begin{eqnarray}
\frac{\partial \psi}{\partial t} = \frac{c}{z} \psi \frac{\partial z}{\partial t}\\
\frac{\partial \psi}{\partial r_x} = \frac{c}{z} \psi \frac{\partial z}{\partial r_x}\\
\frac{\partial \psi}{\partial r_y} = \frac{c}{z} \psi \frac{\partial z}{\partial r_y}\\
\frac{\partial^2 \psi}{\partial r_x^2} = \frac{c}{z} \psi \frac{\partial^2 z}{\partial r_x^2}+
\frac{c^2}{z^2} \psi (\frac{\partial z}{\partial r_x})^2 -\frac{c}{z^2} \psi (\frac{\partial z}{\partial r_x})^2\\
\frac{\partial^2 \psi}{\partial r_y^2} = \frac{c}{z} \psi \frac{\partial^2 z}{\partial r_y^2}+
\frac{c^2}{z^2} \psi (\frac{\partial z}{\partial r_y})^2 -\frac{c}{z^2} \psi (\frac{\partial z}{\partial r_y})^2
\end{eqnarray}

Finally, let's substitute Eqs.(15-19) into Schr\"{o}dinger's wave equation
\begin{equation}
i \hbar \frac{\partial \psi}{\partial t} = -\frac{\hbar^2}{2m}[\frac{\partial^2 \psi}{\partial r_x^2} + \frac{\partial^2 \psi}{\partial r_y^2}] + U(r_x,r_y) \psi .
\end{equation}
We obtain
\begin{equation}
i \hbar  \frac{\partial z}{\partial t} + \frac{\hbar^2}{2m}[\frac{\partial^2 z}{\partial r_x^2}+ \frac{\partial^2 z}{\partial r_y^2} + \frac{c-1}{z} ((\frac{\partial z}{\partial r_x})^2 + (\frac{\partial z}{\partial r_y})^2)] -
\frac{z}{c} U(r_x,r_y) = 0.
\end{equation}
Note that the multiplier on potential energy $U$ can be rewritten
\begin{equation}
\frac{z}{c} = \frac{zc^*}{(x^2+y^2)},
\end{equation}
where $c^* = x-iy$.

Eq.(21) can be divided into real (R) and imaginary (I) parts:
\begin{eqnarray}
\frac{\hbar^2}{2m}[\frac{\partial^2 z}{\partial r_x^2}+ \frac{\partial^2 z}{\partial r_y^2} + \frac{x-1}{z} ((\frac{\partial z}{\partial r_x})^2 + (\frac{\partial z}{\partial r_y})^2)] -
\frac{zx}{(x^2+y^2)} U(r_x,r_y) = 0\mbox{   } (R)\\
\hbar \frac{\partial z}{\partial t} + \frac{\hbar^2}{2m}\frac{y}{z} [(\frac{\partial z}{\partial r_x})^2 + (\frac{\partial z}{\partial r_y})^2)] +
\frac{zy}{(x^2+y^2)} U(r_x,r_y) = 0 \mbox{   }(I)
\end{eqnarray}
We now make some simplifications.  Let $x = 1$ and $y = 2$.  We have
$(x^2+y^2) = 5$. This yields as our real and imaginary equations
\begin{eqnarray}
\frac{\partial^2 z}{\partial r_x^2}+ \frac{\partial^2 z}{\partial r_y^2} - 
\frac{2mz}{5\hbar^2} U(r_x,r_y) = 0\mbox{   } (R)\\
\frac{m}{\hbar} \frac{\partial z}{\partial t} + \frac{1}{z}[(\frac{\partial z}{\partial r_x})^2 + (\frac{\partial z}{\partial r_y})^2)] + \frac{2mz}{5\hbar^2} U(r_x,r_y) = 0 \mbox{   }(I)
\end{eqnarray}
Note that Eqs.(25-26) are dictated by Schr\"{o}dinger's wave equation.

For a fixed potential $U(r_x,r_y) = U_f$, simple solutions to the real (R) equation, Eq.(25), are
\begin{equation}
z = e^{\pm \frac{1}{\sqrt 2}(r_x + r_y)\sqrt{ \frac{2m}{5\hbar^2} U_f}} .
\end{equation}

\subsection{The vortex}

The imaginary equation, Eq.(26) is more interesting. Again we consider a fixed potential $U_f$. The term $(\frac{\partial z}{\partial r_x})^2 + (\frac{\partial z}{\partial r_y})^2$ is the equation for a circle whose components are $\frac{\partial z}{\partial r_x}$
and $\frac{\partial z}{\partial r_y}$.  The radius of the circle is determined by $z[\frac{m}{\hbar} \frac{\partial z}{\partial t} +  \frac{2mz}{5\hbar^2} U_f]$. Whether the circle expands or contracts---thereby creating a \emph{spiral}---depends on $\frac{\partial z}{\partial t}$. A spiral is simply a \emph{vortex} restricted to two dimensions.  So, without loss of generality---since we can easily make $z$ a function of three location co-ordinates---we shall refer to a spiral as a vortex. Thus we have that the imaginary part of the Schr\"{o}dinger equation yields for the wave function $\psi = z^c$ a vortex solution.  Waves can be vortices also.  Recall in this connection Eqs.(10-11), which show that changes in $c$ do not materially change the nature of the underlying wave.  So our use of specific values for $x$ and $y$ in $c = x+iy$ does not change the nature of this conclusion. In the spirit of 'there be dragons here' written on medieval maps, we can say 'there be vortexes here'.

One simple solution to the imaginary (I) equation, Eq.(26), is
\begin{equation}
z = e^{k(r_x+r_y)-3k^2\frac{\hbar}{m} t}
\end{equation}
where
\begin{equation}
k = \sqrt{ \frac{2m}{5\hbar^2} U_f} .
\end{equation}

We can see the motion of the vortex by solving for $u$ and $v$ in
\begin{equation}
\psi = u+iv = z^c = e^{k(r_x+r_y)-3k^2 \frac{\hbar}{m}t + i2k(r_x+r_y)-6ik^2\frac{\hbar}{m} t} .
\end{equation}
Hence 
\begin{eqnarray}
u = e^{k(r_x+r_y)-3k^2\frac{\hbar}{m} t }cos(2k(r_x+r_y)-6k^2 \frac{\hbar}{m}t) \\
v = e^{k(r_x+r_y)-3k^2\frac{\hbar}{m} t }sin(2k(r_x+r_y)-6k^2 \frac{\hbar}{m}t).
\end{eqnarray}
In the (u,v) plane for $\psi$, the initial radius  $\sqrt{(u^2+v^2)}$ of the vortex at $t = 0$ is $e^{k(r_x+r_y)}$.

What is the meaning of $(\frac{\partial z}{\partial r_x})^2 + (\frac{\partial z}{\partial r_y})^2$? We find
$\frac{1}{z}[(\frac{\partial z}{\partial r_x})^2 + (\frac{\partial z}{\partial r_y})^2] = 2k^2 z$. So the derivatives
of $z$ with respect to the position co-ordinates describe a circle with radius $kz\sqrt{2}$. Now, since $z$ is
changing with time $t$, that means the radius of the circle is changing. Assuming $r_x+r_y > 0$, the radius shrinks to $k\sqrt{2}$ at time $t = \frac{m}{\hbar} \frac{r_x+r_y}{3k}$ where $z = 1$ and the wave function $\psi = z^c$ collapses into $\psi = u = 1$.  An alternative solution to the imaginary equation, Eq.(26), is $z = e^{-k(r_x+r_y)-3k^2 \frac{\hbar}{m}t}$, in which case $z$ and the radius of the circle shrink to a point as $t \rightarrow \infty$ and $z\rightarrow 0$, and the wave function $\psi = z^c$ collapses into the origin.  Thus, for an ensemble of vortex waves, their collapse produces a binary sequence of 0s and 1s.

We have assumed $r_x+r_y > 0$.  If $r_x+r_y < 0$, we simply multiply $r_x+r_y$ by -1 and apply the other solution.
So, without loss of generality, we will assume $r_x+r_y > 0$.

If a vortex collapses into 0, we will call it a \emph{0-vortex}; if a vortex collapses into 1, we will call it a \emph{1-vortex}. The 0-vortices and 1-vortices occur as solution pairs to the imaginary (I) Schr\"{o}dinger equation.  If they are produced in equal quantities, but have different times-to-live, there will be an asymmetry in the observed types of vortices, and hence in their collapsed remainders---more 0s than 1s, or vice-versa. Since 0-vortices live
longer than 1-vortices, there are initially more 1-vortices that have collapsed into 1s, and thus more 1s than 0s.  However, since 0-vortices are longer lived, we presume their number increases relative to 1-vortices, until eventually
the number of 0-vortices collapsing to 0 at any time is equal to the number of 1-vortices collapsing to 1. Thus the
number of 0s and 1s would occur in equal proportions, but there would be more 0-vortices than 1-vortices.

The wave function $\psi$ can be normalized to unity.  We will calculate normalizing constants $A_0$ and $A_1$ for 0-vortices and 1-vortices, respectively..  For 0-vortices, the value of the integral
\begin{equation}
\int_{0}^{\infty} \psi^*\psi dt = e^{-2k(r_x+r_y)} \int_{0}^{\infty}e^{-6k^2\frac{\hbar}{m} t} dt = \frac{e^{-2k(r_x+r_y)}}{6k^2\frac{\hbar}{m}}.
\end{equation}
So a normalizing constant is $A_0 = e^{k(r_x+r_y)}k\sqrt{6\frac{\hbar}{m}}$.  For 1-vortices, the value of the integral
\begin{equation}
\int_{0}^{\frac{m}{\hbar}\frac{r_x+r_y}{3k}} \psi^*\psi dt = e^{2k(r_x+r_y)} \int_{0}^{\frac{m}{\hbar}\frac{r_x+r_y}{3k}}e^{-6k^2 \frac{\hbar}{m}t} dt =
 \frac{e^{2k(r_x+r_y)}}{6k^2\frac{\hbar}{m}} \int_0^{2k(r_x+r_y)} e^{-y} dy =
 \frac{e^{2k(r_x+r_y)}-1}{6k^2\frac{\hbar}{m}}.
\end{equation}
So a normalizing constant  is $A_1 = k\sqrt{6\frac{\hbar}{m}}(e^{2k(r_x+r_y)}-1)^{-\frac{1}{2}}$.  From the ratio
of the squared normalzing constants, $[A_0/A_1]^2$, we expect the ratio of 0-vortices to 1-vortices to be
$[e^{4k(r_x+r_y)}-e^{2k(r_x+r_y)}]$, because this would equalize probabilities in the production of 0s and 1s.

Now consider again the gradient compoents of $z$.  For 1-vortices, both differential operators $\frac{\partial}{\partial r_x}$ and $\frac{\partial}{\partial r_y}$ have the same eigenvalue $k$.  Thus the three-dimensional map $(\frac{\partial z}{\partial r_x},\frac{\partial z}{\partial r_y},z)$ begins at the lower boundary point $(k,k,1)$ and forms a straight line away from the origin given by $(kz,kz,z)$ for $z>1$.  For 0-vortices, the eigenvalues of the differential operators are both $-k$, so the same map begins at the upper boundary point (on $z$) $(-k,-k,1)$ and forms a straight line into the
origin given by $(-kz,-kz,z)$ for $z<1$.  There is a one-to-one mapping between these two lines given by
$(kz,kz,z) \rightarrow (-k/z,-kz,1/z)$ for $z>1$.  As $z\rightarrow \infty$, $1/z \rightarrow 0$.  We can think
of constructing these two line seqments as follows.  We start with a 45-degree line, a ray, in the First Quadrant
of the plane $(\frac{\partial z}{\partial r_x},\frac{\partial z}{\partial r_y})$, and snip it at $(k,k)$.  We
reflect the shorter snipped piece around the origin to $(-k,-k)$ in the Third Quadrant.  The two line segments form the basis for the vortex pairs.  However, if we square the co-ordinates, and map $(k^2 z^2, k^2 z^2,z^2)$, the two line segments are reunited into a continuous ray in the First Quadrant begining at the origin.  Thus the two vortex solutions are hidden in $\frac{\partial^2 z}{\partial r_x^2}+\frac{\partial^2 z}{\partial r_y^2}$ by a quadratic uniting of the two line segments which are in fact disconneted in the $(\frac{\partial z}{\partial r_x},\frac{\partial z}{\partial r_y})$ plane.

\subsection{Energy levels in the vortex}

From Eqs.(31-32) the multiplier on time $t$ is
\begin{equation}
6k^2 \frac{\hbar}{m} = \frac{E}{\hbar}
\end{equation}
and thus the energy $E$ is
\begin{equation}
E = \frac{12}{5}U_f.
\end{equation}
The total energy $E$ is $\frac{12}{5}$ the potential energy $U_f$.  This value is a function of our choices for
$x$ and $y$.  

For quantized energy levels, we can replace $U_f$ by the potential $U(E)$.  Let $\{E_j\}$ be a set of real numbers
(eigenvalues) with $E_0<E_1<...<E_j<E$, where we define $j$ by the unit step function $\lambda(\cdot)$, namely, $j+1 = \sum_{i=0}^{\infty} \lambda(E-E_i)$.  Then we define $U(E)$ as
\begin{eqnarray}
U(E) = \frac{5}{12} E_0, j = 0\\
U(E) = \frac{5}{12}[\sum_{i=0}^{\infty} \lambda (E-E_i)E_i-\sum_{i=0}^{j-1}E_i], j>0.
\end{eqnarray}
Note that $\sum_{i=0}^{\infty} \lambda (E-E_i)E_i$ is simply the sum of eigenvalues less than $E$. Eq.(36) is replaced by $E = E_j$. Then a solution for the 1-vortices is, from Eq.(28),
\begin{equation}
z = e^{k(r_x+r_y)-3k^2 t}
\end{equation}
where
\begin{equation}
k = \sqrt{ \frac{2m}{5\hbar^2} U(E)} .
\end{equation}
The potential $U(E)$ is the largest energy eigenvalue $E_j$ permitted by $E$ (multiplied by a scale factor $\frac{5}{12}$), so the solution for $\psi = z^c$ is quantized with angular frequency $\omega = \frac{E_j}{\hbar}$.  

Upon a transition from $E_{j-1}$ to $E_j$, the change in $k$ is
\begin{equation}
\triangle k = \sqrt{ \frac{m}{6\hbar^2}} [\sqrt{E_j}-\sqrt{E_{j-1}}],
\end{equation}
so that $k$ follows the jump process
\begin{equation}
dk = \sqrt{ \frac{m}{6\hbar^2}} \sqrt{dE}.
\end{equation}
Similar reasoning applies to 0-vortices. For a fixed $k$, the time evolution is a smooth function of $t$, but
jumps as energy $\triangle E$ is released or absorbed according to the step function $U(E)$.

\subsection{Conclusion}
This letter has demonstrated that the ordinary linear Schr\"{o}dinger wave equation generates vortex solutions for
the class of wave functions represented by $\psi = z^c$, where $z>0$ is real, and $c = x+iy$ is complex.
Recently the author generalized Heisenberg's uncertainty principle to apply to stable amplitudes $s_{\alpha,\beta}$ \cite{JOG}. Heisenberg considered only the Gaussian case where $\alpha = 2$. However, for stable amplitudes with
$\alpha \rightarrow 0$, quantum uncertainty occurs at a macroscopic level.  This in itself suggests that applying
Schr\"{o}dinger's equation(s) to macroscopic domains may be legitimate.  Moreover, when one finds that vortices are
an intrinsic part of the Schr\"{o}dinger equation, at least in certain circumstances, then the occurrence of vortices
at all levels of the Universe is suggestive of a fractal structure, where correspondences are obtained by scaling and dimensional adjustment.  (The Hausdorf dimension of a stable process $s_{\alpha,\beta}$ is $\alpha$.) In this case we would not need Bohr's correspondence principle of a transition from the quantum to the classical level:  the fractal structure itself would \emph{be} the correspondence principle. All of this, of course, must be subjected to observation
and experiment.

In this vane, Nottale's quantization of planetary motion isn't so far-fetched.  Of course Nottale \cite{LN} has argued for a scale-relative universe, but depletes much of the force of his thesis by also arguing that $\alpha$ must equal 2. 
Another Gaussian interpretation, coupling the Schr\"{o}dinger equation to the Poisson equation and leading to a lognormal galaxy distribution, is described in \cite{PC}.  

Neto \emph{et. al.} \cite{Neto} use a Schr\"{o}dinger-type diffusion equation to calculate planetary mean distances,
the results of which agree with all known planetary orbits and the asteroid belt in the solar system, with three empty slots.  In deriving their result they substitute for $\hbar$ the much larger parameter $g^* = 3.6 \times 10^{42} J\cdot s$.  This suggests the domain of applicability of Schr\"{o}dinger's equation is a matter of scaling ('as below, so above').

Volovik \cite{GV} examines vortices in superconductors, superfluids, and low-density atomic Bose-Einstein condensates,
without making any reference to the Schr\"{o}dinger wave equation. However, a nonlinear Schr\"{o}dinger equation has
been explicitly offered as a model of superfluidity \cite{RB}. This is further indication of the universal applicability
of the Schr\"{o}dinger equation.

\end{document}